\documentclass[aps,pre,twocolumn,floatfix,showpacs]{revtex4}
\usepackage{amsmath}  
\usepackage{amssymb}  
\usepackage{graphicx}

\newcommand{\vv}[1]{{\rm {\bf {#1}}}}  
\newcommand{\vh}[1]{{\rm {\bf\hat{{#1}}}}}  
\newcommand{\lb}{{\langle}}
\newcommand{\rb}{{\rangle}}

\begin{document}

\title{Collision statistics of driven granular materials}

\author{Daniel L. Blair and A. Kudrolli}

\affiliation{Department of Physics, Clark University, Worcester, MA
01610, USA}

\date{\today}

\begin{abstract} 
We present an experimental investigation of the statistical properties
of spherical granular particles on an inclined plane that are excited
by an oscillating side-wall. The data is obtained by high-speed
imaging and particle tracking techniques.  We identify all
particles in the system and link their positions to form trajectories
over long times. Thus, we identify particle collisions to measure the
effective coefficient of restitution and find a broad distribution of
values for the same impact angles.  We find that the energy
inelasticity can take on values greater than one, which implies that
the rotational degrees play an important role in energy transfer.  We
also measure the distance and the time between collision events in
order to directly determine the distribution of path lengths and the
free times.  These distributions are shown to deviate from expected
theoretical forms for elastic spheres, demonstrating the inherent
clustering in this system.  We describe the data with a two-parameter
fitting function and use it to calculated the mean free path and
collision time.  We find that the ratio of these values is consistent
with the average velocity.  The velocity distribution are observed to
be strongly non-Gaussian and do not demonstrate any apparent universal
behavior.  We report the scaling of the second moment, which
corresponds to the granular temperature, and higher order moments as 
a function of distance from the driving wall.
Additionally, we measure long time correlation functions in both space
and in the velocities to probe diffusion in a dissipative gas.
\end{abstract} 

\maketitle
\section{\label{sec:intro}Introduction}
 
Granular material represent a type of matter not well defined by
conventional means. Although each granular particle  is obviously
solid, an assemblage of these particles show distinctly  non-solid
behavior when subjected to external forces~\cite{jaeger96}. In the
rapid flow regime,  the interaction between the grains is collisional
and the system resembles  a dense {\em granular gas}. Indeed, the
kinetic theory for dense gases formulated by Chapman and
Enskog~\cite{chap} have been modified to include the dissipative
nature of the collisions ~\cite{jenkins83,haff83}.   However, a number
of approximations have to be made in any calculation that can be only
validated by experiments.  Furthermore, even if  key assumptions such
as equipartition breakdown ~\cite{mcnamara92,grossman97,kadanoff}, it
is important to have a measure of the failure to guide further
development.

Energy has to be constantly supplied from an external source to
observe a steady state in granular gas systems. Therefore, model
experiments consist of granular particles inside a container where
energy is continuously injected at a
side-wall~\cite{warr94,kudrolli97,olafsen98}.  Thus gradients are
present in experimental granular systems, which implies that care must
be taken when comparing results to non-equilibrium kinetic
theory~\cite{noije98,gtz,pug98}.   With advances in high speed image
acquisition, it is now possible to obtain positions of particles
several times between collisions. However, particle positions and
velocities can be obtained accurately only in two dimensions by direct
imaging thus forcing certain constraints on the geometry of the system.

One of the first such experiments to investigate velocity distribution
functions (VDFs) utilized an apparatus in which particles are vibrated
vertically inside a narrow transparent
box~\cite{warr94,warr95_1,warr95_2}.  Maxwellian statistics were
reported for the vertical and horizontal velocity components of the
particles parallel to the plane of the transparent
side-walls. Additional interactions in these systems arise due to
collisions between particles and the side-walls~\cite{warr95_1}.
Following this work, Wildman {\em et al.}~\cite{wildman} were able to
do long time particle tracking to measure diffusion constants by
interpreting mean square displacement data over a very broad range of
density.  More recently, in a similar apparatus, Rouyer and
Menon~\cite{rouyer00} report that their VDFs have a universal form
that can be parameterized by a single variable, the granular
temperature. A different method of energy injection utilizes large
flat container that is vibrated vertically to excite a sub mono-layer
of particles~\cite{olafsen98,olafsen99,losert99}.  The velocity of the
particles in the horizontal plane are measured and are found to follow
a non-Gaussian distribution. However, the impact of the velocity
gradient in the vertical direction on the observed distributions are
not taken in to account because these components cannot be measured.

Our experiment is a variation of the vertically vibrated apparatus.
Spherical particles are constrained to roll on an inclined two
dimensional surface.  This geometry allows for a direct investigation
of the interplay between energy injected at the side-wall and the
dissipation through inelastic collisions.   In addition, the 
inclination reduces the effects of gravity, therefore minimizing shock
waves.  This system has been used to demonstrate clustering and
collapse when the inter-particle collision frequency is much greater
than particle-driving wall collision
frequency~\cite{kudrolli97}. Recent works have explored a full range
VDFs, from very near Gaussian behavior to highly non-Gaussian
distribution functions, as well as velocity
correlations~\cite{kudrolli00,blair01}.

In addition to analytical techniques and experiments, several groups
have utilized computer simulations of inelastic hard spheres with both
Molecular
Dynamics~\cite{gtz,netp1,moon01,netp2,vanzon02,pug_02,brey02} and
Direct Simulation
Monte-Carlo~\cite{pug_pre,brey,balda01,morgado,pug_pre01} techniques
to investigate the statistical  properties of granular gases.  Using
DSMC simulations, Baldassarri {\em et al.}~\cite{pug_pre} have found
velocity and density distributions that are qualitatively similar to
our previous experimental results~\cite{blair01}.  Recent work by Brey
and Ruiz-Montero~\cite{brey02} investigate how the second and fourth
moments of the VDFs scale as a function of distance from the driving
wall, which until now, have not been experimentally tested.

In this paper, we report on the statistical properties of a gas of
inelastic particles constrained to two-dimensions.  An inclined
geometry  reduces the gravitational acceleration acting on each
particle which results in lower mean velocities.   The combination of
slow dynamics and high speed imaging allows us to accurately identify
the particle trajectories and collision events.   By using velocities
before and after a collision event, we measure the normal coefficient
of restitution.  We find that these quantities are found to be broadly
distributed for the same impact parameters.  By calculating the
distance and time between collision events we measure the
distributions of free paths and times.  We find that these
distributions do not follow the result found from kinetic theory.  The
path and time distributions have an overpopulation of short distance
and time bins, demonstrating the inherent clustering present in
granular gases.  We propose an empirical form that captures the
distributions, which is then used to calculate the mean free path and free
time as a function of density.  The particle trajectories are also
used to measure the mean square displacement, velocity
auto-correlation, and diffusion rates.  The distribution of particle
velocities are measured with a variation in density of an order of
magnitude and show distinctly non-Gaussian behavior with no apparent
universal form.  We compare our results to recent experiments, as well
as theoretical and simulation treatments of equivalent systems.

The paper has the following structure.  In Section~\ref{sec:methods}
we present the experimental apparatus and imaging methods.
Section~\ref{sec:char} provides the overall system characteristics
such as the density distributions and  coefficients of restitution and
inelasticities.  We then present our analysis
of the trajectories of the particles in Sec.~\ref{sec:results}.  
Finally, in Sec.~\ref{sec:disc} we summarize our results in the 
context of granular kinetic theory and simulations.
 
\begin{figure}[tb]
\includegraphics[width=0.45\textwidth]{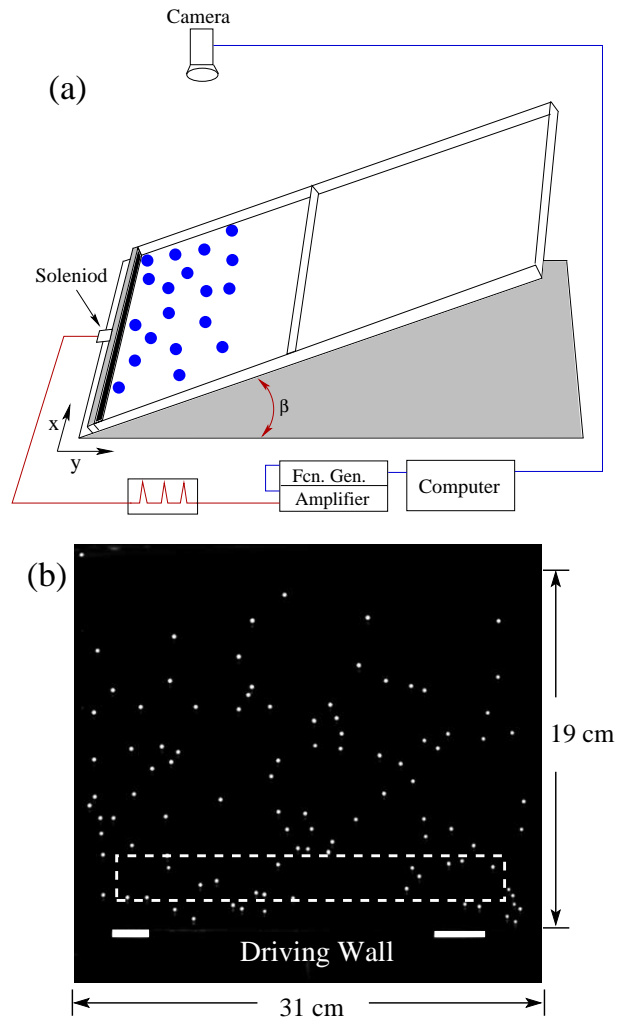}
\caption{ \label{fig:apparatus}(a) Schematic diagram of the
experimental setup.  The inclined plane is a smooth glass surface, the
side-walls and driving wall are stainless steel so that the
particle-boundary collisions approximate those between particles.  The
driving is produced by a solenoid connected to the lowest side-wall.
The angle of inclination $\beta$, can be varied from $\beta=0-8$
degrees, the values of $\beta$ we have chosen are $2^{\circ}\pm
0.1^{\circ}$ and $4^{\circ}\pm 0.1^{\circ}$.   (b) An image of the
system taken from above.  The bottom right corner is considered the
origin of our coordinate system $(0,0)$.  The white bars allow us to
track the position of the driving wall.}
\end{figure}

\section{\label{sec:methods}Experimental Methods}

\begin{figure}[tb]
\includegraphics[width=0.45\textwidth]{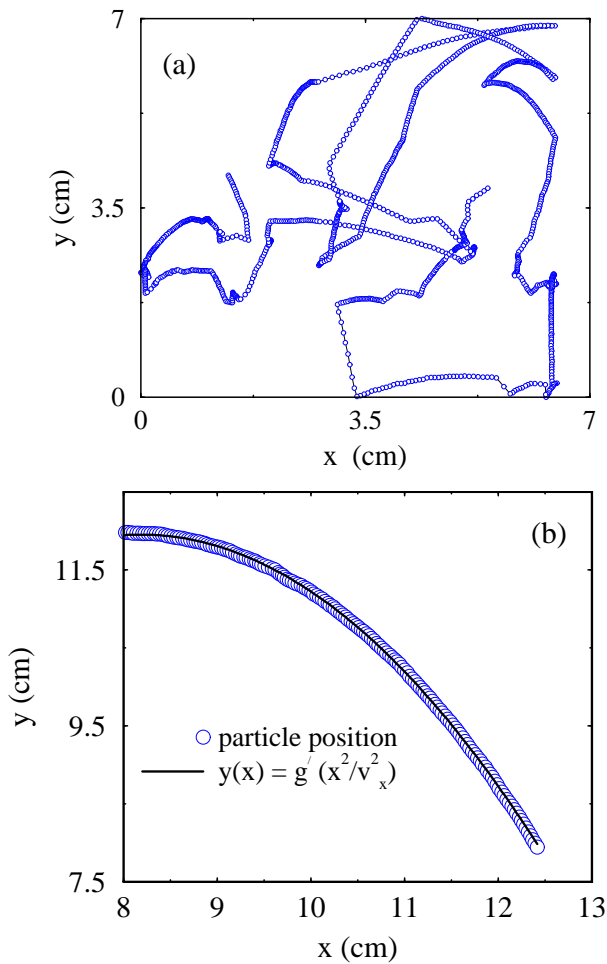}
\caption{\label{fig:traj}(a) Linked particle positions over 1365 time
steps ($N_l=3$).  We can determine particle collision events with a
high degree of accuracy from trajectories such as this. (b) The
parabolic path of a particle. We use  fixed values for $v_x$ and $g =
980$ cm s$^{-2}$ to measure $g' = \frac{5}{7}g \sin \beta$.  The fit
gives $\beta = 2.2^{\circ}$, the deviation from the measured value of
$\beta$ is $9\%$.}
\end{figure}

The experimental configuration [Fig.~\ref{fig:apparatus}], consists of
a $100$~$\sigma$ x 60~$\sigma$ glass plane that is inclined at an
angle $\beta$ with the horizontal.  The particles are stainless steel
with diameter $\sigma =3.175$~mm and a high degree of sphericity
($\delta \sigma / \sigma = 10^{-4}$).  The number of particles,
measured in number of mono-layers $N_l$ across the driving wall, is
varied between $N_l=\,$ 1--5 in steps of one layer, ({\em viz.}  from
$N_p = 100-500$ in steps of 100, where $N_p$ is the number of
particles).  The energy source is an oscillating side-wall, driven by
a solenoid, that is located as shown in Fig.~\ref{fig:apparatus}(a).
The driving signal is a 10~Hz pulse with a velocity during each pulse
of $\sim 40$~cm s$^{-1}$.  The driving frequency and amplitude were
chosen to ensure that no phase dependence on the center of mass is
observed (at frequencies below 2~Hz the particle positions are
phase-locked with the driving). The signal is produced with a computer
interfaced Aglient Technologies 33120-A wave-form generator  that is
and subsequently amplified by an HP 6824A Amplifier. The inclination
of the plane can be varied between $\beta = 0^{\circ}- 8^{\circ}$, for
our experiments the angle was fixed at $\beta=2^{\circ}\pm
0.1^{\circ}$ or $4^{\circ}\pm 0.1^{\circ}$.  In the extreme case of
$\beta \ll 1^{\circ}$, the particles essentially cease to interact
with the energy source and cluster at the side opposite of the driving.

The particles are imaged using a  Kodak MotionCorder SR1000 high speed
digital camera.  We measure the positions of all particles contained
in the apparatus for 1365 frames at $250$ frames per second at full
spatial resolution of $512\times 480$ pixels.  These digital images
are then transfered to a computer and analyzed using a centroid method
that allows us to resolve each particle to sub-pixel accuracy.  After
each particle is located the positions are then connected in time to
form continuous trajectories for 5.46~s.  Our coordinate system is
such that the $x,y$ axes are parallel and perpendicular to the driving
respectively [see Fig.~\ref{fig:apparatus}].  A typical particle
trajectory is shown in Fig.~\ref{fig:traj}(a).  Multiple collision
events can be distinguished with nearly straight paths between each
event.  A particle that freely rolls on the inclined plane  will
follow a parabolic trajectory [see Fig.~\ref{fig:traj}(b)].  The
particle trajectory is given by
\begin{equation}
y(x) = \frac{5}{7}\frac{x^2 }{2v^{2}_{x}}g \sin(\beta) ,
\label{eq:parab}
\end{equation}
where, $g$ is the acceleration due to gravity, $v_x$ is measured from
the width of the parabola, and the $\frac{5}{7}$ factor is due to the
moment of inertia for a solid sphere.

\begin{figure}[t]
\includegraphics[width=0.45\textwidth]{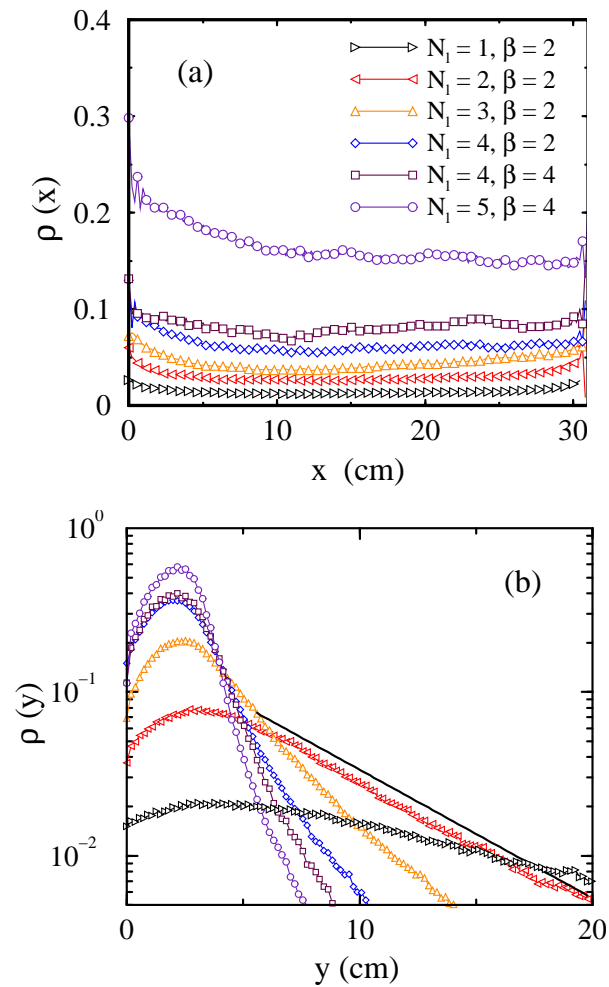}
\caption{\label{fig:dens}(a) The density $\rho(x)$ versus $x$ for all
$N_l$.   The obvious clustering due to inelastic collisions at  the
side-walls is demonstrated here.  Also, as $N_l$ is increased the
system becomes more inhomogeneous in across the cell.  This effect is
most likely due to the onset of clustering instabilities that have
been recently discussed \cite{meerson_02_1,meerson_02_2,soto_02}. (b)
The aerial density plots $\rho(y)$ for each $N_l$ and $\beta$ on a
log-linear plot. The density $\phi$ is measured in a particular area
by integrating $\rho(y)$ over that region of interest.  The total are
under each curve corresponds to the average area fraction for that
particular $N_l$.  The solid line shows an exponential fit over the
tail of the distribution of $N_l=200$.  However, we will demonstrate
that the isothermal atmosphere is not obeyed for any
density.}
\end{figure}

\begin{table}[b]
\begin{ruledtabular}
\begin{tabular}{lccc}
$N_l$&  $\beta$&  $\phi$&  $N_p$ \\ \hline \hline  1& 2.0& 0.022& 100
\\  2& 2.0& 0.068& 200 \\  3& 2.0& 0.138& 300 \\ 4& 2.0& 0.191& 400 \\
4& 4.0& 0.302& 400 \\ 5& 4.0& 0.581& 500 \\
\end{tabular}
\caption{\label{table1} Experimental values of the number of layers
$N_l$, the angle of inclination $\beta$, and the resulting measured
value of $\phi$. $N_p$ the number of particles in the system is given
for clarity.}
\end{ruledtabular}
\end{table}

\section{\label{sec:char}System Characteristics}

\subsection{\label{sub:dens}Density Distributions}

The results presented will be given in terms of the number of single
layers across the cell, $N_l$ and the angle of inclination $\beta$,
which the determine the area fraction $\phi$ [see Table~\ref{table1}].
We measure $\phi$ by defining  a region of interest (ROI) that is
centered about the peak in $\rho(y)$, [Fig~\ref{fig:dens}(b)] whose
extent in the $y$-direction is limited to $\pm 10\% \text{ of }
\rho(y)$. The ROI scheme excludes all particles that are within
3$\sigma$  of the side walls to ensure that clustering due to the
side-walls does not affect our results [see Fig.~\ref{fig:dens}(a)].
The over-plotted box in Fig.~\ref{fig:apparatus}(b) demonstrates the
ROI definition for $\phi =0.13$.  A more stringent division of the
system in the $y$-direction will be used when the behavior of the
temperature, pressure and kurtosis are discussed in
Sec~\ref{sec:results}.   The form of the  density in the $y$-direction
is similar to that found in Refs.~\cite{luding94,warr95_1} however, we find
that the form of the tails of $\rho(y)$ at higher values of
$N_l\,\beta $ deviate from Boltzmann distribution.  This implies that
the law of isothermal atmospheres breaks down for granular systems as
we shall also see when we discuss the scaling of the granular
temperature in Sec.~\ref{sub:vel}.


\subsection{Particle Collisions}

We identify collision events from the trajectories by using the
following algorithm.  Velocities are constructed as finite differences
${\vv{v}}_j = \frac{\Delta \vv{x}}{\Delta t}$, where $\Delta \vv{x} =
\vv{x}(t_j) - \vv{x}(t_i)$  and the subscripts $i,j$ represent
positions separated by the time difference $\Delta t = 4\,$ms.  All
velocity vectors are compared sequentially to find direction changes
given by
\begin{equation}
\psi = \cos^{-1}(\vh{v}_i \cdot \vh{v}_j),
\label{gam_dot}
\end{equation}
where $\vh{v} = \vv{v}/|\vv{v}|$ the unit vector of the calculated
velocity.  If  $20^{\circ} \le \psi \le 180^{o}$ the proximity of  all
particles at the same time instant is checked.  If a particle is found
within a radius $\sigma +\Delta\sigma$, whose velocity also satisfy
Eq.\ (\ref{gam_dot}) it  is considered as a candidate for a collision.
To assure that re-collisions are not occurring, we maintain a record
of the identity of the previous collision partner.  We then ensure that those
particles can re-collide if and only if the partner particle  has
undergone a collision with yet a third particle.  If particles pass
these requirements then a collision has occurred.  To extend the
algorithm to include collisions with the boundary walls we first check 
if Eq.\ (\ref{gam_dot}) is satisfied.  We then check if the particle's
center is within $\sigma + \Delta\sigma$ of a boundary and it's
velocity component perpendicular to the wall is reversed.

\subsection{Coefficient of Restitution}

The loss of energy in a collision is given by the coefficient of
restitution. Two particles that undergo an inelastic collision with a
relative velocity between particles $\vv{v}_{12} = \vv{v}_1 -
\vv{v}_2$, will obey the reflection law  $\vv{v}_{12}^* \cdot
\hat{\sigma} = - \alpha \, \vv{v}_{12} \cdot \hat{\sigma}$, where
$\alpha$ is the normal component of the restitution coefficient and
$\hat{\sigma}$ is the unit vector connecting the centers of the
particles.

\begin{figure}[tb]
\includegraphics[width=0.5\textwidth]{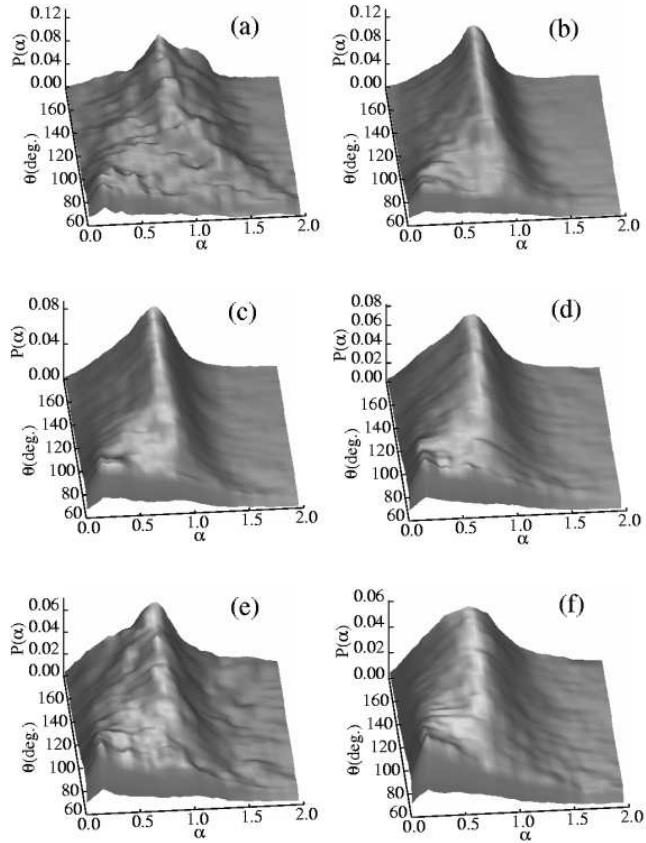}
\caption{\label{fig:restI}The distribution of the normal component of
restitution $\alpha$ versus $60^{\circ} \le \theta \le 180^{\circ}$,
the relative angle of incidence between particle velocities.  (a)
$N_l=1\,,\beta=2.0$, (b) $N_l=2\,,\beta=2.0$, (c) $N_l=3\,,\beta=2.0$,
(d) $N_l=4\,,\beta=2.0$, (e) $N_l=4\,,\beta=4.0$, (f) $N_l=5 \,
\beta=4.0$.  The value of the z-axis for each graph is the probability
of a collision giving a value of $\alpha$ in a range of $\theta +
\Delta\theta$, where $\Delta\theta = 2^{\circ}$ }
\end{figure}
\begin{figure}[t]
\includegraphics[width=0.45\textwidth]{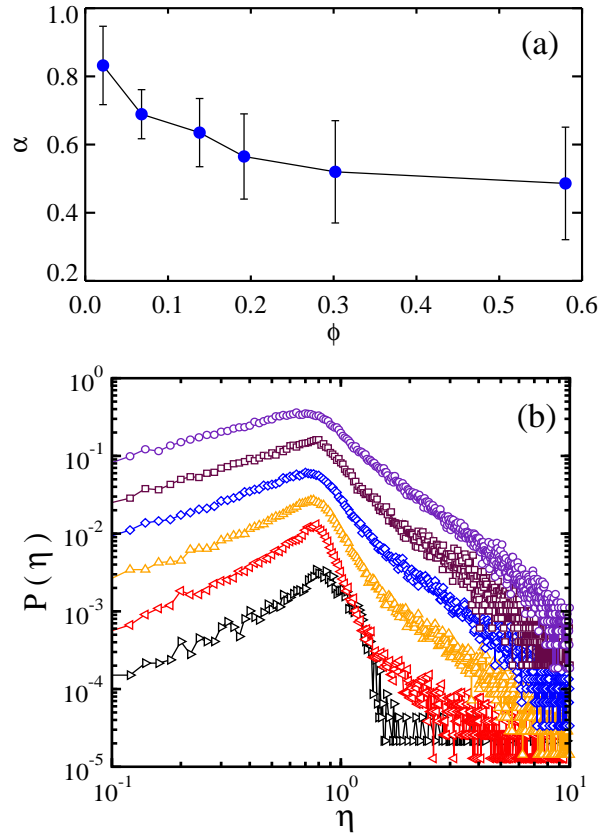}
\caption{\label{fig:mean_rest}(a) The mean value of the distributions
of $\alpha$ shown in Fig.~\ref{fig:restI} averaged over $60^{\circ}\ge
\theta \ge 180^{\circ}$, as a function of the average covering
fraction $\phi$.  The bars indicate the spread in the
distribution. (b) The distribution of energy inelasticities given by
Eq.\ (\ref{eq:inelas}) for $(\rhd) N_l=1\,,\beta=2.0$, $(\lhd)
N_l=2\,,\beta=2.0$, $(\vartriangle) N_l=3\,,\beta=2.0$, $(\diamond)
N_l=4\,,\beta=2.0$, $(\Box) N_l=4\,,\beta=4.0$, $(\circ) N_l=5 \,
\beta=4.0$.  Each distribution is shifted vertically for clarity.}
\end{figure}

Having an efficient method for collision identification, we are able
to measure the relative velocities of two particles before and after
collision events.   The coefficient of normal restitution during a
binary collision is given by
\begin{equation} 
\alpha  = -\frac{(\bar{\vv{v}}^{*}_{12}\cdot
\hat{\sigma})}{(\bar{\vv{v}}_{12} \cdot \hat{\sigma})},
\end{equation}
where the over-bar denotes average over three pre/post-collisional
velocities measured in the ROI described above [see
Fig.~\ref{fig:apparatus}(b)].  The angle between the relative
velocities of two colliding particles is given by
\begin{equation}
\theta = \cos^{-1}(\vv{\bar{v}}_{12}\cdot \vv{\bar{v}}^{*}_{12}).
\label{eq:theta}
\end{equation}
Thus we can characterize the coefficient of restitution as a function
of $\theta$.  The probability distributions $P(\alpha)$ for
$60^{\circ} \le \theta \le 180^{\circ}$ for each $N_l\,\beta$, are
shown in Fig.~\ref{fig:restI}(a--f).  Data for $\theta < 60^{\circ}$
suffers from a lack of statistics and therefore is not included.  Each
graph represents the probability of the inelasticity having a value
$\alpha$ for a range of $\theta +\Delta\theta$, where
$\Delta\theta=2^{\circ}$.  $P(\alpha)$ follows a very broad
distribution of values over all $\theta$, and have a decreasing mean
value as function of $\phi$ [see Fig.~\ref{fig:mean_rest}(a)].  Thus
we find that the coefficient of restitution can have a broad
distribution of values for the same impact angle.

We also measured the energy loss due to a collision as a function of
$N_l\,\beta$.  The ratio of the magnitudes of the relative velocities
before and after a collision,
\begin{equation}
\eta = \frac{|\bar{\vv{v}}^{*}_{12}|}{|\bar{\vv{v}}_{12}|},
\label{eq:inelas}
\end{equation}
determines the {\em energy} restitution coefficient, ($\eta^2 =
\alpha^2$ if all $\theta$ are averaged).
Figure~\ref{fig:mean_rest}(b) shows the distributions of measured
values of $\eta$ shifted for clarity.   We find that a peak exists at
a value that is consistent with $\alpha^2$.   Furthermore there exists
a power law tail for values of $\eta > 1$, which has been interpreted
as a {\em random inelasticity}~\cite{barrat}.  The appearance of a
tail at high $\eta$ implies that the rotational degrees of freedom are
actively transferring energy to translational motion during a
collision.

\section{\label{sec:results}Results}

\subsection{Distributions of Paths and Times}

\begin{figure}[t]
\includegraphics[width=0.5\textwidth]{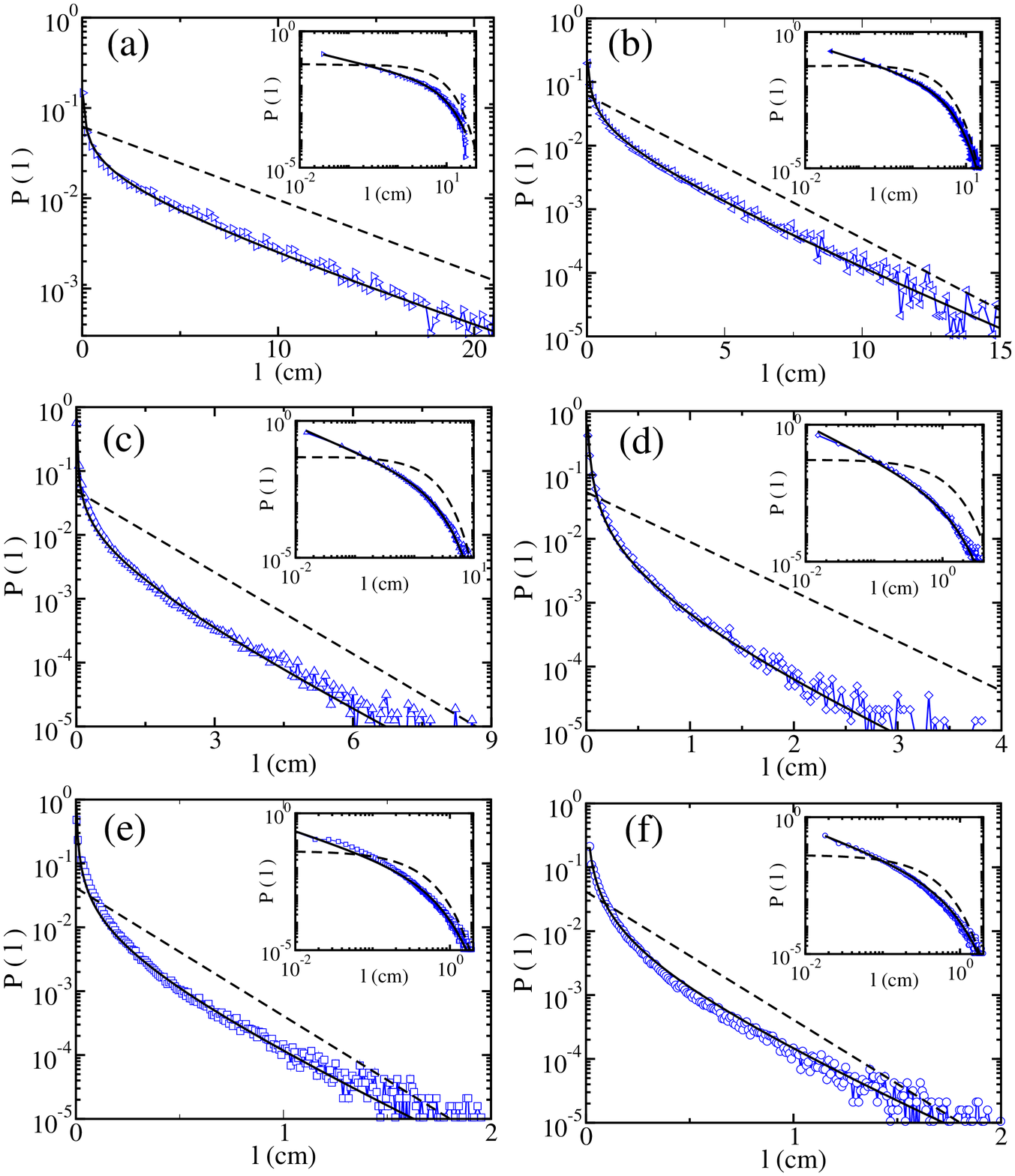}
\caption{\label{fig:path_dist} The probability distributions of path
lengths $P(l)$ versus $l$, on a log-linear scale, and
{\em inset} log-log scale (a)  $N_l=1\,,\beta=2.0$, (b)
$N_l=2\,,\beta=2.0$, (c) $N_l=3\,,\beta=2.0$, (d) $N_l=4\,,\beta=2.0$,
(e) $N_l=4\,,\beta=4.0$, (f) $N_l=5\,,\beta=4.0$. The dashed line
shows the theoretical form given by Eq.\ (\ref{eq:poflam}) derived for
elastic particles, and the solid  line is an empirical fit given by
Eq.\ (\ref{eq:landtau}a).  Table~\ref{table2} shows the fit
parameters. }
\end{figure}

We measure the distribution of paths lengths from the the geometric
distance between collision events defined in our ROI at each $N_l\,
\beta$ [Fig.~\ref{fig:path_dist}(a--f)]. By basic kinetic theory
arguments~\cite{hecht}, the distribution of path lengths for an
elastic hard-sphere gas (and by a similar treatment the distribution
of free times) is given by,
\begin{equation}
\label{eq:poflam}
P(l) =   (2\sqrt{2}\, \,\,\phi)\,\, e^{\,-2\sqrt{2}\,\phi\, l}.
\end{equation}
The distribution therefore should follow a simple exponential form
depending only on the density.  However it is clear from the dashed
lines in Fig.~\ref{fig:path_dist}(a--f) that the simple form given by
Eq.\ (\ref{eq:poflam}) does not describe the behavior over all $l$.

\begin{figure}[t]
\includegraphics[width=0.5\textwidth]{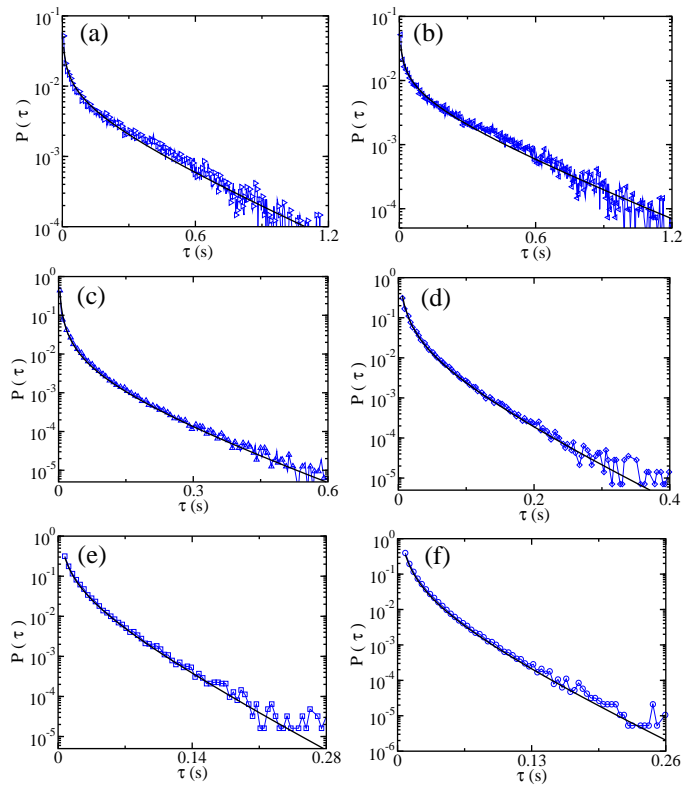}
\caption{\label{fig:time_dist} The probability distributions of free
times $P(\tau)$ versus $\tau$, on a log-linear scale
(a) $N_l=1\,,\beta=2.0$, (b) $N_l=2\,,\beta=2.0$, (c)
$N_l=3\,,\beta=2.0$, (d) $N_l=4\,,\beta=2.0$, (e) $N_l=4\,,\beta=4.0$,
(f) $N_l=5\,,\beta=4.0$. The solid line is a fit given by Eq.\
(\ref{eq:landtau}b)}
\end{figure}

The distributions of times between collisions $P(\tau)$
[Fig.~\ref{fig:time_dist}(a--f)] is also measured and shows similar
behavior to that of the path length distributions, that is an
overpopulation of the short time bins.  This should be expected from
the simple relationship between the displacement and the time.
However, it is noteworthy to mention that the ratio of $l/\tau$ versus
the path length $l$,  is not a constant over all values of $l$,
implying that the average speed of the system depends on the distance
or time between collisions.

Elastic hard spheres will have a mean free path that is simply
$\bar{l} = \bar{v} \bar{\tau}$, where $\bar{v}$ and
$\bar{\tau}$ are the average speed and collision time respectively.
Also, the mean free path can be derived directly from the distribution
of path lengths, $\bar{l} = \int_0^{\infty} l\,P(l)\, dl$, where
$P(l)$ is given by Eq.~\ref{eq:poflam}.  Grossman {\em et
al.}~\cite{grossman97} have interpolated how the mean free path for a
granular system should be modified to account for higher collision
rates due to increased density.  Although the interpolation gives a
qualitatively accurate correction for passing between the high and low
density limits, the actual distribution of path lengths has not been
measured or calculated for a granular gas.

We have found an empirical form that well describes the measured
distributions of path lengths and free times,
\begin{subequations}
\label{eq:landtau}
\begin{eqnarray}
P(l) &=& a \, (l)^{-b}\, e^{-c \,l}\label{eq:pofl}\\ P(\tau) &=& a \,
(\tau)^{-b}\, e^{-c\, \tau}\label{eq:poft}
\end{eqnarray}
\end{subequations}
where $a,b,c$ for the path lengths and free times are shown in
Table~\ref{table2} for all $N_l\,\beta$. This form appears to capture
both the short $l$ and $\tau$ power-law behavior.  In the dilute case
the form returns to the theoretical prediction for larger path lengths.

\begin{figure}[t]
\includegraphics[width=0.45\textwidth]{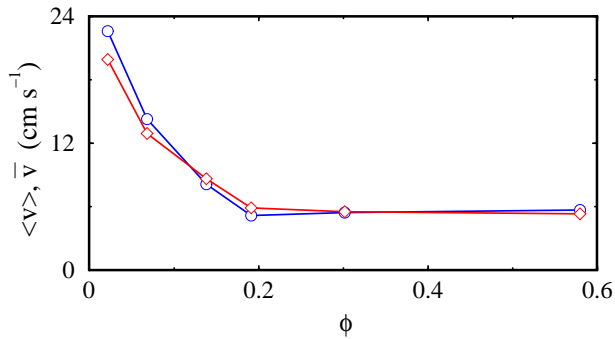}
\caption{ \label{fig:ave_vel} The average speed measured for each
$N_l\,\beta$. ($\circ$) $\bar{v}$ obtained from Eq.~(\ref{eq:avevel}),
and ($\diamond$) $\lb v \rb$ measured from the mean of the speed
distribution.  The two independent measures give similar values over
the entire density range.}
\end{figure}

From the distribution of path lengths and free times, we calculate the
mean free path and time by utilizing the fitting form and it's
parameters.  The ratio of the mean free path to the mean collision
time should determine the average speed $\bar{v}$ in the ROI
where the distributions are measured.  We have taken the ratios of the
integrated distributions,
\begin{equation}
\label{eq:avevel}
\bar{v} = \frac{\bar{l}}{\bar{\tau}} = \frac{\int_0^{\infty}
l\,P(l)\, dl}{\int_0^{\infty} \tau P(\tau)\, d\tau},
\end{equation}
and compared that to the average of the speed distribution $\lb
v_{x,y}\rb$ in the same ROI.  Figure~\ref{fig:ave_vel} the both
measurements for all $N_l \,\beta$.  The agreement is within $10 \%$ over
the entire range of $N_l\, \beta$ indicating that the proposed forms
in Eqs.~(\ref{eq:landtau}a,b) quantitatively capture the behavior of
the distributions.

\begin{table}[b]
\begin{ruledtabular}
\begin{tabular}{lcccc}
$N_l$& $\beta$&  $a(a)$&  $b(b)$&  $c(c)$ \\ 
\hline \hline  
1& 2& 0.031(0.0027)& 0.428(0.511)& 0.154(2.969) \\  
2& 2& 0.025(0.0025)& 0.603(0.665)& 0.393(6.024) \\ 
3& 2& 0.008(0.0003)& 0.932(1.203)& 0.742(8.306) \\
4& 2& 0.003(0.0008)& 1.258(1.209)& 1.489(16.91) \\
4& 4& 0.002(0.0002)& 0.970(1.043)& 3.171(26.45) \\ 
5& 4& 0.002(0.0005)& 1.096(1.384)& 2.887(28.33) \\
\end{tabular}
\caption{\label{table2} Fitting parameters for
Eqs.~(\ref{eq:landtau}a,b).  The values are arranged $a(a)$ for
$P(l) (P(\tau))$, respectively.  The $(\cdots)$ correspond to the
values for $P(\tau)$}
\end{ruledtabular}
\end{table}

\subsection{Velocity Autocorrelation}

\begin{figure}[tb]
\includegraphics[width=0.45\textwidth]{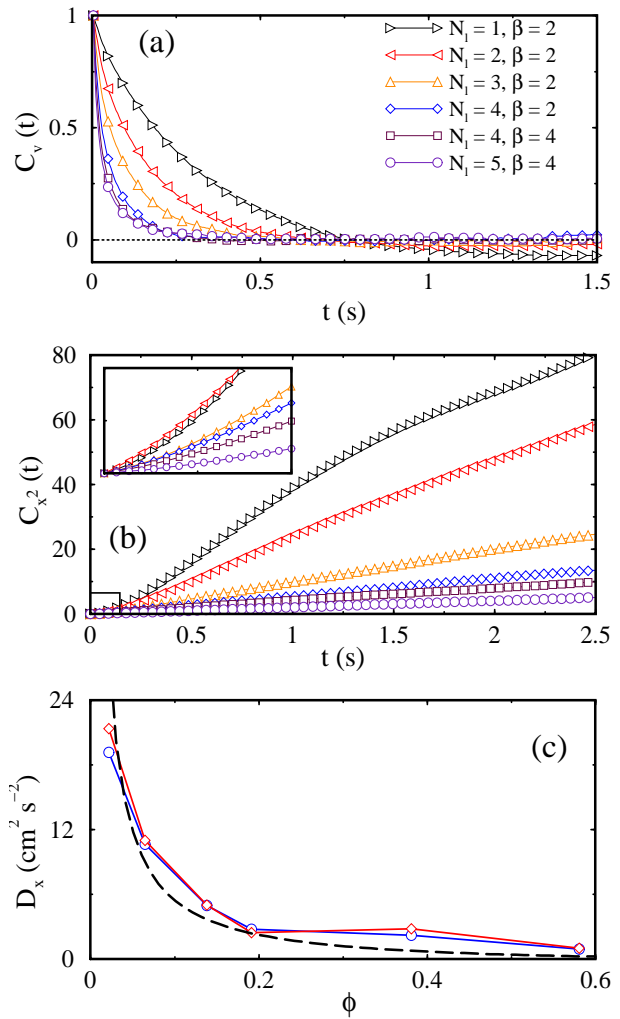}
\caption{(a) The velocity auto correlation function $C_{v}(t)$. (b)
The mean square displacement in the $x$-direction for each $N_l\,\beta
$ for $t=0-2.5\,$s. {\em Inset:} The short time behavior indicated by
the box in the main figure. (c) The diffusion constants calculated for
each $\phi$, where ($\circ$) corresponds to the numerical integration of
$C_v$ from (a), and ($\diamond$) corresponds to the least squares fit of
$C_{x^2}$ from (b).  The dashed line shows the kinetic theory result
for a fixed temperature, which is given by the average measured value
over this range of $\phi$.}
\label{fig:auto_corr}
\end{figure}

The velocity autocorrelation function (VAF) is computed for the $x$
components of the velocities within an ROI by using the
following~\cite{allen}:
\begin{equation}
C_{v}(t) = \frac{1}{N_p N_s t_{max}
}\sum_{i,j=0}^{N_p,N_s}\sum_{\Delta t=1}^{t_{max}}  \vv{v}_{ij}(t_o)
\cdot \vv{v}_{ij}(t_o + \Delta t)
\label{eq:vel_auto}
\end{equation}
where, $N_s$ is the total number of data sets, $N_p$ is the number of
particles and $t_{max}$ is the total number of time origins.
Figure~\ref{fig:auto_corr}(a), shows the measured values of the VAF
normalized by $\lb \vv{v}(0)^2\rb$ in our system.

In simulations of hard sphere fluids, Alder and
Wainwright~\cite{alder67} first found that the form for the VAF was
strongly depended on the density of the system.  For  very low
densities the the characteristic form of the correlation function was
given simply by
\begin{equation}
C_{v}(t) = \lb \vv{v}(0)^2\rb e^{\,-t/\tau_E},
\label{eq:auto_fit}
\end{equation}
where $\tau_E$ is the Enskog collision time.  If the density of the
system is increased however, the form of Eq.\ (\ref{eq:auto_fit})
breaks down and $C_{v}(t)$  can become negative with long range tails
due to the caging of particles by their neighbors.  We find, that the
{\em lowest} density case becomes, and remains negatively correlated
after the decay from $\lb \vv{v}(0)^2\rb$
[Fig~\ref{fig:auto_corr}(a)].  This appears to be in contradiction
with Ref.~\cite{alder67}  but is due to a finite size effect. That is,
the particles are interacting frequently with the side-walls at low
densities, which reverse the sign of velocity vectors,
thus leading to the observed anti-correlation.  The predominance of
the sidewall interactions are screened for the intermediate densities
due to the increased number of particle-particle collisions, therefore
no anti-correlations are observed.

\subsection{Mean Square Displacement}

To determine the mean square displacement of the $x$-component of the
particle positions [Fig~\ref{fig:auto_corr}(b)], by using the
sequential time zeros for each trajectory~\cite{allen} given by
\begin{equation}
C_{x^2}(t) = \frac{1}{N_p N_s t_{max}
}\sum_{i,j=0}^{N_p,N_s}\sum_{\Delta t=1}^{t_{max}}  |\vv{x}_{ij}(t_o)
- \vv{x}_{ij}(t_o + \Delta t)|^2.
\label{eq:msd}
\end{equation}
Where $N_s$ is the total number of data sets, $N_p$ is the number of
particles and $t_{max}$ is the total number of time origins. For these
measurements, the ROI is allowed to increases in size along the
$y$-direction as $N_l$ decreases.  We have chosen to make this
increase to ensure that particles at low $N_l$ have had the
opportunity to undergo a collision while under consideration.

The long time behavior of $C_{x^2}$ for each $N_l\,\beta$
[Fig.~\ref{fig:auto_corr}(b)], displays linear dependence on time,
indicating diffusive behavior.  However for $N_l =1$, $C_{x^2}$
clearly shows a crossing from one linear regime to another, which may
be a possible indication of finite system size for low density. For
short times, [Fig.~\ref{fig:auto_corr}(b){\em Inset}] the behavior is
ballistic as indicated by the quadratic increase of $C_{x^2}$ in time.
As $N_l$ is increased, the range of the ballistic regime dramatically
decreases indicating a decrease in the Enskog collision time $\tau_E$.
The ballistic and diffusive regimes are consistent with what is
expected for kinetic theory of elastic, finite-sized particles.

\subsection{Self-Diffusion}

The self diffusion constant $D$, can be determined for a system of
particles by either evaluating the time integral of the velocity
autocorrelation function,
\begin{equation} 
D = \int_0^{\infty} C_{v} (t) dt,
\label{eq:diff_auto}
\end{equation}
or using the relationship between the mean square displacement of the
particles and time over long times,
\begin{equation}
D = \lim_{t\to\infty} \,\frac{1}{2\,d\,t} C_{x^2}(t),
\label{eq:diff_rsquare}
\end{equation}
where $d$ is spatial dimension. From kinetic theory~\cite{hecht}, the
diffusion constant of a two dimensional gas is calculated as:
\begin{equation}
\label{eq:ens_diff}
D = \frac{\sigma}{8 \phi g(\sigma)}\left(\frac{\pi T}{m}\right)^{1/2}
\end{equation}
where, $g(\sigma)$ is the radial correlation function at
contact~\cite{henderson} given by,
\begin{equation}
\label{gofr}
g(r=\sigma) = \frac{16 - 7\phi}{16(1-\phi)^2}.
\end{equation}
By numerically integrating the curves in Fig.~\ref{fig:auto_corr}(a),
and performing a least squares fit to the data in
Fig.~\ref{fig:auto_corr}(b) after the ballistic regime, we obtain the
self-diffusion constant [see Fig.~\ref{fig:auto_corr}(c)].  We find
that the values for the self-diffusion from
Eqs.~(\ref{eq:diff_auto},\ref{eq:diff_rsquare}) are self consistent.
The solid line in Fig.~\ref{fig:auto_corr}(c) shows the form of Eq.\
(\ref{eq:ens_diff}) with the temperature $T$ given by the
granular temperature. The granular temperature is defined by
\begin{equation}
T_{x,y}= \frac{1}{2} m [ \lb v^2_x \rb + \lb v^2_y \rb],
\label{eq:gran_temp}
\end{equation} 
where, $m$ is the mass of the particles and $\lb \cdots \rb$ denote
averages over the component distributions [see Sec.~\ref{sub:vel}].
The other constants in Eq.~\ref{eq:ens_diff} are determined from
system parameters.  The theory for the diffusion of elastic particles, 
given by Eq.~(\ref{eq:ens_diff}) closely matches our results all
$\phi$.  Thus we show that the effects of inelasticity on the
self-diffusion are small. 

\subsection{\label{sub:vel}Velocity Distributions}

\begin{figure}[t]
\includegraphics[width=0.46\textwidth]{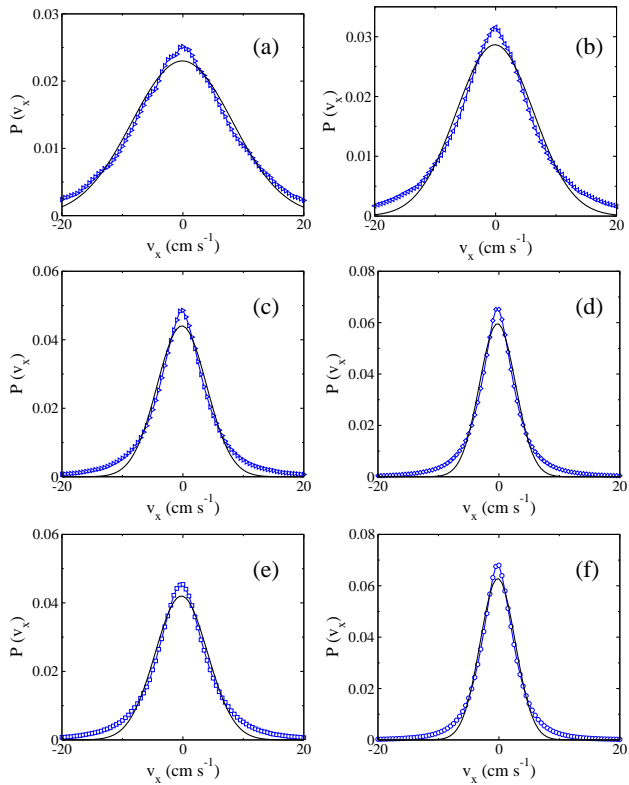}
\caption{\label{fig:velx_lin}The velocity distribution functions
$P(v_x)$ versus $v_x$ on a linear-linear scale.  (a)
$N_l=1\,,\beta=2.0$, (b) $N_l=2\,,\beta=2.0$, (c) $N_l=3\,,\beta=2.0$,
(d) $N_l=4\,,\beta=2.0$, (e) $N_l=4\,,\beta=4.0$, (f)
$N_l=5\,,\beta=4.0$. The solid curves are a least squares fit to a
Gaussian form given by Eq.\ (\ref{boltz}). Note that the deviation
from a Gaussian distribution extends all the way to the lowest
velocity bins.  Each distribution correspond to $\sim 2\times 10^6$
unique velocities that are found within the ROI defined in the text.}
\end{figure}

\begin{figure}[t]
\includegraphics[width=0.46\textwidth]{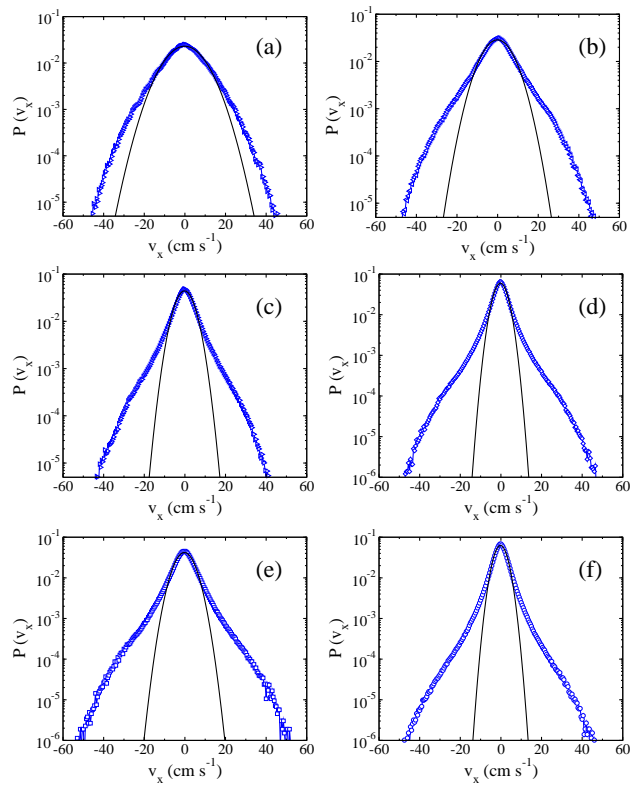}
\caption{\label{fig:velx_log}The velocity distribution functions
$P(v_x)$ versus $v_x$ on a log-linear scale.(a) $N_l=1\,,\beta=2.0$,
(b) $N_l=2\,,\beta=2.0$, (c) $N_l=3\,,\beta=2.0$, (d)
$N_l=4\,,\beta=2.0$, (e) $N_l=4\,,\beta=4.0$, (f)
$N_l=5\,,\beta=4.0$. The solid curves are a least squares fit to a
Gaussian form given by Eq.\ (\ref{boltz}). Here the apparent deviation
in the tails of the distribution functions are present.}
\end{figure}

\begin{figure}[t]
\includegraphics[width=0.46\textwidth]{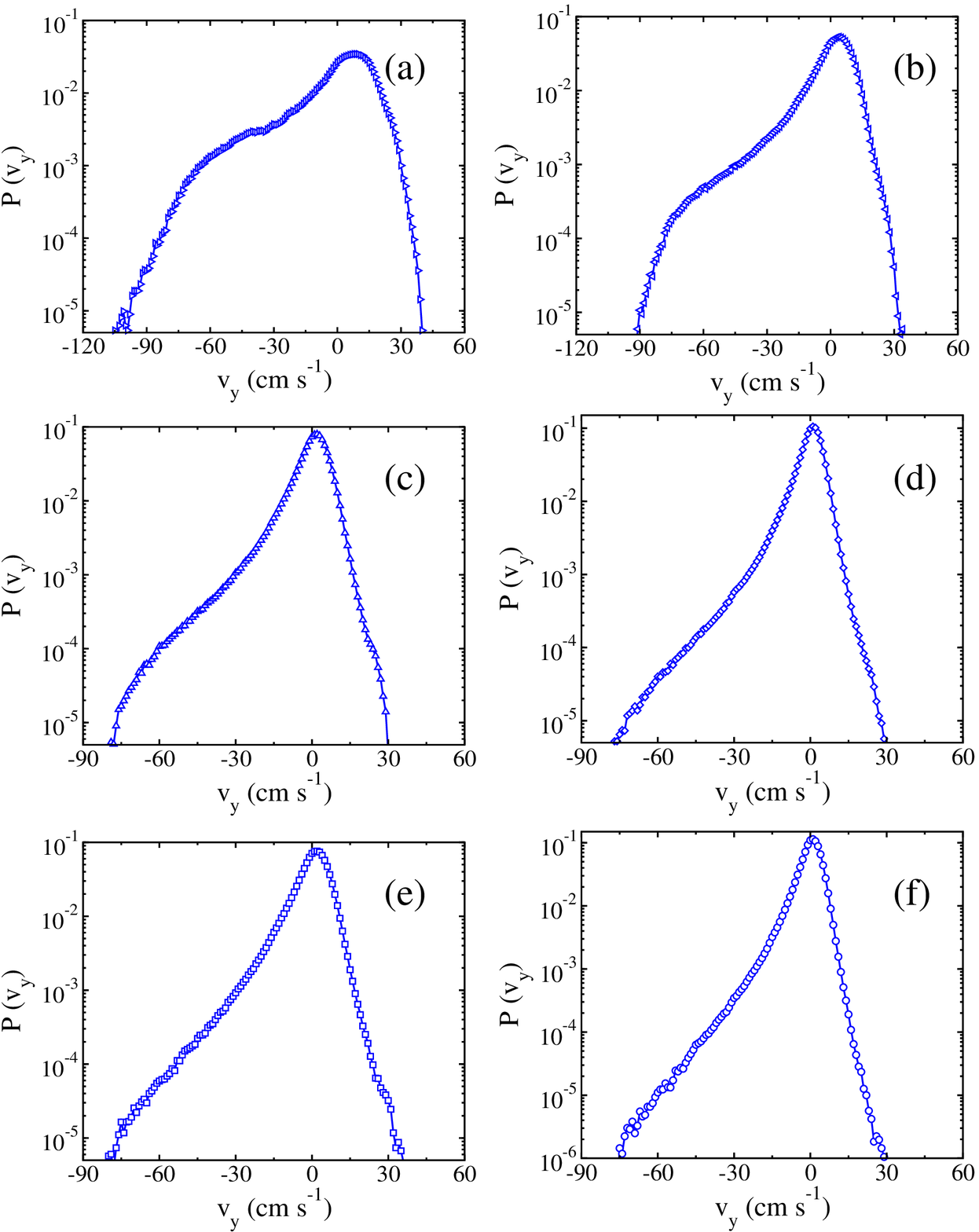}
\caption{\label{fig:vely_log}The velocity distribution functions
$P(v_y)$ versus $v_y$ on a log-linear scale.(a) $N_l=1\,,\beta=2.0$,
(b) $N_l=2\,,\beta=2.0$, (c) $N_l=3\,,\beta=2.0$, (d)
$N_l=4\,,\beta=2.0$, (e) $N_l=4\,,\beta=4.0$, (f)
$N_l=5\,,\beta=4.0$. The large skewness in the distributions for the
negative values of $v_y$ is due primarily to the driving from the
bottom wall. Particles that are moving in the $-y$ direction are
leaving the moving wall.}
\end{figure}

The distribution of the $x$- and $y$-components of the particle
velocities are plotted in
Figs.~\ref{fig:velx_lin}--\ref{fig:vely_log}(a--f).  The distributions
correspond to velocities that are measured within a region of
interest.  The ROI is defined by making a narrow slice across the
$y$-direction that is centered upon the peak in $\rho(y)$ while
excluding particles lying within a distance of $ 3 \sigma$ from the
side walls.  We utilize this ROI to ensure that large gradients in
$\rho(y)$, and the clustering produced by the side-walls, do not
affect the measured  VDFs.  Each distribution correspond to $\sim
2\times 10^6$ unique velocities that are found within our ROI.

The velocities of elastic particles follow a distribution given by the
Maxwell-Boltzmann form, 
\begin{equation}
P(\vv{v}) = (2 \pi k_BT)^{-d/2} e^{-\vv{v}^2 / 2k_BT}
\label{boltz}
\end{equation}
where, $d$ is the dimensionality of the system and $T$ is the
temperature of the heat bath that the system is in contact with.
Hence, if a system of particles is at equilibrium, its temperature
determined by the width of the distribution of particle velocities.
Equation~(\ref{boltz}) is fit to the data for the $x$-components of
the velocities, and is shown on both linear and logarithmic scales
[Fig.~\ref{fig:velx_lin},\ref{fig:velx_log}].   We observe that the
form given by Eq.~(\ref{boltz}) displays deviations both at low and
high velocities.  The distributions of velocities are normally
displayed in a log-linear fashion to accentuate the tails of the VDF,
however this suppresses the deviations at low velocities.  By plotting
the distributions on a linear scale we display the more statistically
significant deviations from Eq.~(\ref{boltz}).

In a recent experimental work~\cite{rouyer00}, a two dimensional
collection of particles is driven into a steady state.  Using
analysis techniques that are similar to ours, the authors proposed 
a governing form for the VDF given by
\begin{equation}
R(\vv{v}) = A\, e^{-B\, | \vv{v}_x /T_x|^{-1.5}},
\label{eq:rouyer}
\end{equation}
where $A$ and $B$ are constants and $T_x$ is the $x$-component of the
granular temperature defined in Eq.~(\ref{eq:gran_temp}). They
claim to have seen a {\em universal} VDF which they parameterized by a
single value, regardless of the system density or the value of the
inelasticity of the particles.   From our
VDFs, whose corresponding densities range over an order of magnitude
and where the average inelasticity varies by nearly a factor of two, 
we cannot find any single parameter fit that describes the overall form.

The VDFs for the $y$ components, $P(\vv{v}_y)$ versus $\vv{v}_y$  for
each $N_l\,\beta$ in our ROI are also measured [see
Fig.~\ref{fig:vely_log}(a--f)]. The VDFs are highly skewed by the
asymmetry in the driving against the direction of gravity.  To
identify the effects that the asymmetry in $P(\vv{v}_y)$ has upon
$P(\vv{v}_x)$, we have separated the $\vv{v}_x$ distributions by the
sign of $\vv{v}_y$, {\em i.e.}  $P(\vv{v}_x|+\vv{v}_y;-\vv{v}_y)$.  We
have found that the form for these conditional distributions are not
affected by the sign of $\vv{v}_y$, however we do note that their
widths differ, with $\lb \vv{v}_x\rb_{+\vv{v}_y} < \lb
\vv{v}_x\rb_{-\vv{v}_y}$.

\begin{figure}[t]
\includegraphics[width=0.45\textwidth]{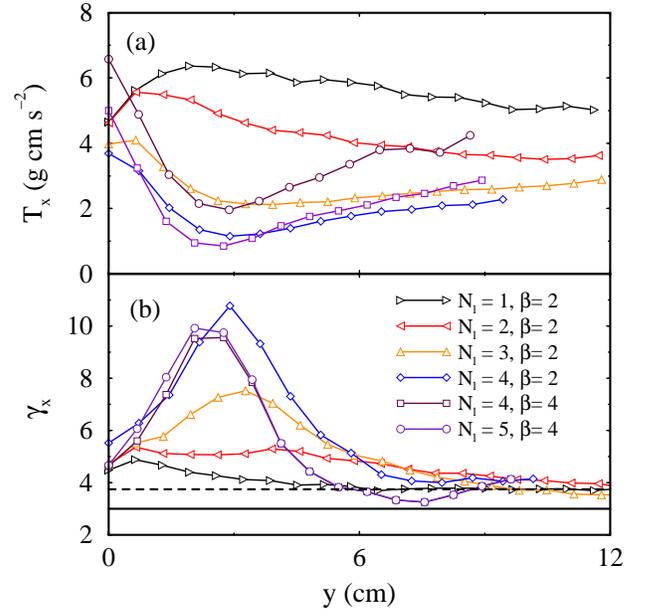}
\caption{\label{fig:temp}(a) The granular temperature $\frac{1}{2}m\lb
\vv{v}_x^2\rb$ as a function of distance from the driving wall for
each $N_l\,\beta$.  If the isothermal atmosphere condition was
satisfied these would be constant values for all $y$ above the peak in
$\rho(y)$.  For values of $N_l > 2$ the temperatures follow a
non-monotonic form that has a distinct minimum.  (b) The kurtosis,
$\gamma_x$ measured from $P(\vv{v}_x)$ as a function of the distance
from the driving.  The values given by a Gaussian (solid line) and the
form proposed in Eq.\ (\ref{eq:rouyer}) (dashed) are only attained
very far from the energy source. }
\end{figure}

We also measure the $x$-component of the granular temperature $T_x$
[see Eq.~(\ref{eq:gran_temp})], to probe the scaling behavior of the
velocity distributions.  Figure~\ref{fig:temp}(a) shows the measured
granular temperature as a function of distance from the driving wall.
At low densities ($N_l \le 2$), $T_x(y)$ initially increases and then
decays.  In contrast, for ($N_l \ge 3$), $T_x(y)$ has a distinct
minimum.  We note that $T_x(y)$ never reaches a constant value and the
minimum (maximum) does not correspond to the peak in $\rho(y)$.

To further show the non-universality of the VDFs, we plot the kurtosis
as a function of distance from the driving wall. The kurtosis is
obtained by the following:
\begin{equation}
\gamma = \frac{\lb v_{x}^4\rb}{\lb v_{x}^2\rb^2}.
\label{eq:kurt}
\end{equation}
If the velocity distribution is a  Gaussian then $\gamma=3$, shown by
the solid line in Figure~\ref{fig:temp}(b), and if the distribution is
given by Eq.\ (\ref{eq:rouyer}) then $\gamma=3.576$.    We find that
the the measured values for $\gamma$ exceed the value for a Gaussian
and also vary as a function of distance from the driving.   This
analysis is consistent with our previous results~\cite{blair01} and
recent MD simulations of Brey and Ruiz-Montero~\cite{brey02} that
closely mimic our experiment.

\subsection{Equation of State}

\begin{figure}[t]
\includegraphics[width=0.45\textwidth]{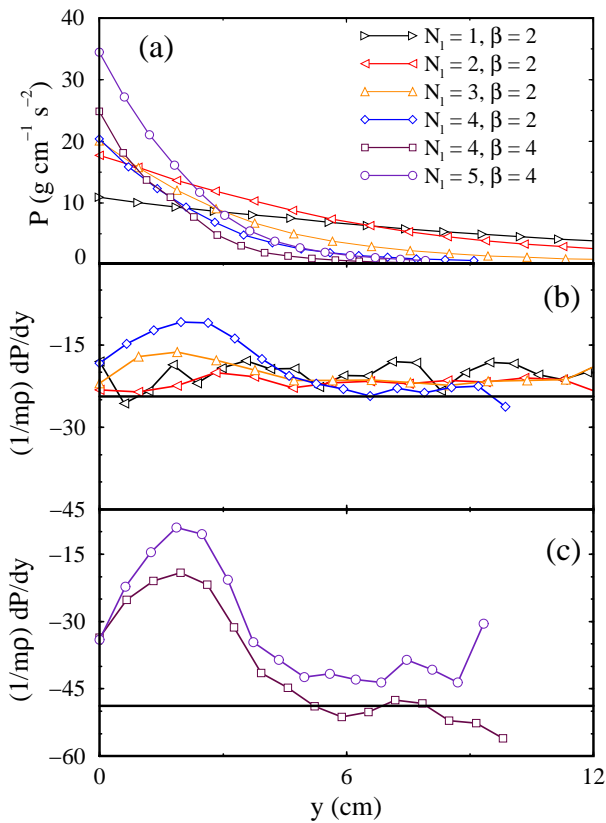}
\caption{\label{fig:p_dp} (a) The pressure $P(y)=\rho T$ as a function
of $y$, the distance from the driving wall. (b) The ratio of the mass
density to the granular pressure force $\frac{1}{m\rho}\frac{dP}{dy}$
as a function of distance from the driving wall.  The solid line
corresponds to $g'$ for $\beta = 2^{\circ}$ (c) The same as (b) for
$\beta = 4^{\circ}$.  The obvious deviation for regions of high
density show a breakdown of the simple treatment of the granular
equation of state in regions of high density.}
\end{figure}

The equation of state for ideal gases relates the pressure to the
temperature and the density,
\begin{equation}
P = n k_B T,
\label{eq:press}
\end{equation}
where $n$ is the number density and $k_B$ is Boltzmann's constant.  If
we assume that kinetic theory is valid for a granular gas, we can
immediately relate the average squared speed, $\lb v^2 \rb$ of the
particles to the temperature,
\begin{equation}
m\lb v^2\rb = k_B T,
\label{eq:equipart}
\end{equation}
for each degree of freedom.   Figure~\ref{fig:p_dp}(a) shows the
pressure, $P=\frac{nm}{2}\lb v_y^2\rb$ as a function of distance from
the driving wall.  Due to the effects of gravity on the particles, the
density should follow the well known atmospheric law,
\begin{equation}
\label{eq:den}
\rho(y) = \rho_0 \,e^{-mgy/\,T_y},
\end{equation}
which assumes a constant temperature.  We find that the temperature is
not constant for any $N_l\,\beta$ [see Fig.~\ref{fig:temp}].  This is
also consistent with our observations of the density distributions in
Sec.~\ref{sub:dens} where $\rho(y)$ deviates from the form of
Eq.~(\ref{eq:den}).

Momentum balance implies that the gradient of the pressure is related
to $n$ by the following equation:
\begin{equation}
\label{eq:P_force}
\frac{dP}{dy} = - n m g
\end{equation}
where, $m$ is the mass of a particle and $g$ is the acceleration of
gravity.  Due to the non-vanishing gradient of temperature the general
form of the pressure gradient must be taken into account
\begin{equation}
\frac{dP}{dy} = T_y\frac{d n}{dy} + n\frac{dT_y}{dy}.
\label{dpdy}
\end{equation}
We find that Eq.~(\ref{dpdy}) is indistinguishable from the numerical
derivative of $d(n T_y)/dy$.

We have measured the pressure gradient acting on a particle held at a
particular $y$ by evaluating,
\begin{equation}
-\left( \frac{T_y}{nm} \right)\frac{d(nT_y)}{dy} = g',
\label{eq:gravity}
\end{equation}
where $g' = \frac{5}{7}g \sin(\beta)$.  Figure~\ref{fig:p_dp}(b) shows
the LHS of Eq.\ (\ref{eq:gravity}) for $\beta=2$, and
Fig.~\ref{fig:p_dp}(c) for $\beta=4$.  The solid lines correspond to
the values of $g'$ with $\beta = 2,4$ found by utilizing Eq.\
(\ref{eq:parab}) and the data in Fig.~\ref{fig:traj}(b).  We find that
the measured values systematically overestimate the actual values for
$g'$ in the region where the density reaches it's maximum.  Our
interpretation assumes a dilute gas, therefore the deviations near the
peak in $\rho(y)$ are not surprising.

To incorporate the effects due to increased density, we have obtained
the pressure from the interpolated equation of state derived by
Grossman {\em et al.}~\cite{grossman97}:
\begin{equation}
P = n T_y\frac{n_c + n}{n_c - n}
\label{eq:gross}
\end{equation}
where, $n_c$ is the close packing number density. Utilizing
Eq.~\ref{eq:gross}  to calculate the gradient of the pressure
[Eq.~(\ref{eq:P_force})]. However, we find that the disagreement
persists between the LHS of Eq.~(\ref{eq:gravity}) and the measured
value of $g'$ near the peak in $\rho(y)$.

\section{\label{sec:disc}Summary and conclusion}

In this paper we have presented a statistical analysis of an inelastic
gas that is constrained to two dimensions.  Utilizing high speed
digital image processing we perform long time tracking over a broad
range of densities.  Not surprisingly, we observe that the statistical
properties of inelastic gases deviate from expectations of the kinetic
theory for smooth elastic particles.  The most apparent discrepancies
are found in the distribution of free paths and times and the
distribution of particle velocities.

To characterize our system we measure the effective coefficient of
restitution from the relative pre- and post-collision velocities of
particles undergoing binary collisions.  We find that the normal
component of restitution and the energy inelasticity are not
single-valued, but have a distribution of values even for the same
impact parameters.  The mean value of the normal components of
restitution systematically decreases with the system density.  We also
find that the energy inelasticity can take on values greater than
unity, demonstrating a transference of energy from the rotational to
linear  degrees of freedom. In a recent numerical work Barrat and
Trizac have measured the projected one-dimensional coefficient of
restitution Ref.~\cite{barrat,barrat_pre} and the energy inelasticity.
Their interpretation is that the coefficient of restitution and the
energy restitution are random variables that characterize collisions,
consistent with our findings.

The distribution of path lengths and free times are shown to have an
overpopulation of the short distance and time bins.  We have proposed
an empirical form in Equations (\ref{eq:landtau} a,b) which capture
the overall behavior of the observed distributions of paths and times.
By integrating these distribution functions, we are able to measure
the  mean free path and mean time.  The average speed obtained from
the speed distribution and from the mean free path and time are in
close agreement.  Inspired by these finding, Paolotti {\em
et al.}~\cite{pug_02} have reported similar results for the mean free
time in a simulation that mimics our system.

Particle diffusion constants are measured from two independent long
time averaged correlation functions.  The mean square displacement and
the velocity autocorrelation function are calculated.  By then
performing least squares fitting and numerical integration to these
quantities respectively, the self diffusion over a broad range in
density is calculated.  We find that the diffusion constants are
similar to that of a two dimensional gas over this density regime.
Therefore, long-time averaged correlation functions seem to accurately
capture the diffusive properties of granular gases.

We find that the distribution of particle velocities perpendicular to
the direction of driving does not have a universal form, but depends
on both the density and the inelasticity.  In addition, we find a
distinct asymmetry in the VDFs parallel to the driving direction.  We
measure the granular temperature as a function of distance from the
driving source and find non-monotonic behavior.    For low densities,
the granular temperature has a distinct maximum and for high densities
there exists a distinct minimum.  The temperature inversion at higher
densities has recently been described via granular hydrodynamics  by
Ram\'irez and Soto~\cite{ramirez}.  However, the crossover from $T(y)$
having a maximum (for low densities), to $T(y)$ with a minimum at high
densities has not been discussed in any kinetic or hydrodynamic models.

By using kinetic theory and simple hydrodynamics we have tested the
force balance between the gradient of the pressure exerted by a
granular gas on a particle and the force due to gravity
[Eq.~(\ref{eq:P_force})].  We find strong deviations in the regions of
high density.  A simple hydrodynamic form, that describes the behavior
over all densities, is not yet available.

We acknowledge stimulating discussions with H.\ Gould, J.\ Tobochnik,
and A.\ Puglisi. This work was supported by the NSF under Grant
No. DMR-9983659, and by the donors of the Petroleum Research Fund.


\end{document}